\documentclass[preprint]{JHEP3}
\usepackage{epsfig}

\title{Strange meson form factors in holographic QCD}

\author{
H. Z. Sang and Xiao-Hong Wu \\
Institute of Modern Physics,
School of Science,
East China University of Science and Technology,
Meilong Road 130, Shanghai 200237, China \\
E-mail: \email{sanghz@ecust.edu.cn, xhwu@ecust.edu.cn}
}
\preprint{arXiv:1004.xxxx [hep-ph]}

\abstract{
We consider the electromagnetic form factors of
strange vector, axial vector and pseudoscalar mesons
in a holographic QCD model.
We find the charge radius of charged kaon agrees with the experiment,
while the charge radius of charged pion is a little bit smaller than
the experimental value, as obtained in other calculations
in the hard-wall holographic QCD models.
The charge radii of charged rho and $K^\ast$ quantitatively agree with
a recent Dyson-Schwinger equation calculation.
We also present the electric form factors of vector, axial vector
and pseudoscalar mesons in both space-like and time-like regions.
We find the charged kaon form factor is in agreement with
the experiment data.
}

\keywords{AdS/CFT correspondence, hadron physics, electromagnetic form factor}

\begin{document}

\section{Introduction}

Recent progress in string theory, the AdS$/$CFT correspondence,
sheds new light on the strongly coupled gauge theory~\cite{Maldacena:1997re,
Gubser:1998bc,Witten:1998qj}.
This remarkable idea has been applied to study quantum chromodynamics (QCD).
Interesting and successful models have been constructed
to deal with non-perturbative QCD,
in both top-down string approach starting from D-brane configuration
and bottom-up phenomenological approach from low energy hadron dynamics,
for a review, see ref.~\cite{Erdmenger:2007cm}.
Even though the dual model is not real QCD,
essential features of QCD can be captured at least to some degree,
such as confinement and chiral symmetry breaking.
In string theory, top-down models with different D-brane configuration,
$D3/D7$, $D4/D6$ and $D4/D8$ branes
have been proposed to understand QCD and hadron physics~\cite{Karch:2002sh,
Kruczenski:2003be,Kruczenski:2003uq,Sakai:2004cn,Sakai:2005yt}.
From a more phenomenological prospect,
a gauged linear sigma model in AdS$_5$ space, dubbed AdS$/$QCD,
has been constructed to describe low energy
hadron physics~\cite{Erlich:2005qh,DaRold:2005zs}.
This bottom-up approach seems somehow {\it ad hoc} at first glance,
it can capture the generic features of brane model constructions.
Following the AdS$/$CFT correspondence,
chiral symmetry is dual to bulk gauge symmetry,
and the QCD operators $\bar{q}_L \gamma^\mu q_L$, $\bar{q}_R \gamma^\mu q_R$
and bilinear $\bar{q}_R q_L$ are dual to
the bulk gauge fields $L_\mu(x,z)$, $R_\mu(x,z)$ and scalar field $\Phi(x,z)$.
The static hadronic observables,
such as meson mass spectra, decay constants, interaction couplings,
chiral coefficients, have been studied from the holographic QCD model
with the building blocks above.
These hadronic observables are in good agreement with the experiment data
up to the level of $\sim 30\%$.

Hadronic form factors are important to understand the internal structure
of composite particles from underlying QCD dynamics.
The pion form factors have been determined by experiments
at CERN~\cite{Amendolia:1983di}, DESY~\cite{Brauel:1977ra,Ackermann:1977rp}
and JLab~\cite{Tadevosyan:2007yd,Horn:2006tm}.
The static property, charge radius,
and form factor of charged kaon have been measured by UCLA+~\cite{Dally:1980dj}
and CERN NA7 Collaborations~\cite{Amendolia:1986ui}.
There are no experiments capable of measuring the form factors
of vector and axial vector mesons so far.
However, these form factors are important to calculate
nucleon form factors in quark-diquark models,
because the diquark correlator is closely
related to the vector and axial vector form factors~\cite{Cloet:2008re}.
The dynamical hadronic properties,
such as electromagnetic form factors of meson have been studied within
quark model~\cite{Choi:1997iq,Choi:2004ww,deMelo:1997hh},
Dyson-Schwinger equation (DSE) method~\cite{Burden:1995ve,Hawes:1998bz,
Maris:1999bh,Maris:2000sk,Bhagwat:2006pu}
and QCD sum rule~\cite{Braguta:2004kx,Aliev:2004uj}.
Recently, the form factors of pion~\cite{Brodsky:2007hb,Kwee:2007dd,
Grigoryan:2007wn,Kwee:2007nq,Kwee:2008zt,Kim:2008xx}
and rho~\cite{Grigoryan:2007vg,Grigoryan:2007my,BallonBayona:2009ar}
have been calculated in holographic QCD models.

Strange sector mesons have been studied by several groups
in the framework of holographic QCD model.
Interesting results,
such as mass spectra~\cite{Shock:2006qy},
four-point functions needed to investigate the $\Delta I = 1/2$
rule for kaon decays and the $B_K$ parameter~\cite{Hambye:2006av},
$K_{l3}$ transition form factors~\cite{Abidin:2009aj},
and the $U(1)$ problem~\cite{Katz:2007tf}
have been pursued.
In this work, we study the charge radii
and electromagnetic form factors of
strange vector, axial vector and pseudoscalar mesons
in a holographic QCD model.
We find the charge radius and form factor
of charged kaon agree with the experiment data.
The charge radii of charged rho and $K^\ast$ quantitatively agree with
a recent DSE calculation~\cite{Bhagwat:2006pu}.
We also present the electric form factors of vector, axial vector
and pseudoscalar mesons in both space-like and time-like regions.

This paper is organized as follows.
In section 2, we overview the holographic QCD model
and present our notations.
In section 3, we calculate the mass spectra, decay constants,
and electromagnetic form factors of vector, axial and pseudoscalar mesons.
The conclusions are drawn in section 4.

\section{The holographic QCD model}

The three flavor holographic QCD model is defined as
a 5D $SU(3)_L \otimes SU(3)_R$ gauge theory
in the AdS$_5$ space on an interval.
The Lagrangian of the holographic QCD model~\cite{Erlich:2005qh, DaRold:2005zs}
is given by,
\begin{eqnarray}
{\cal L}_{\rm 5} & = & \sqrt{g} M_5 {\rm Tr} \bigg[
 - \frac{1}{4} L_{MN} L^{MN} - \frac{1}{4} R_{MN} R^{MN}
 + \frac{1}{2} (D_M \Phi)^\dagger D^M \Phi
- \frac{1}{2} M_\Phi^2 \Phi^{\dagger} \Phi \bigg] \, ,
\end{eqnarray}
where $M_\Phi^2 = -3/L^2$ from AdS/CFT correspondence
~\cite{Maldacena:1997re,Gubser:1998bc,Witten:1998qj},
$D_M\Phi =\partial_M\Phi +iL_M\Phi -i\Phi R_M$,
$L_M=L_M^a \lambda^a/2$ with $\lambda^a$ being the Gell-Mann matrix,
and $M, N = 0,1,2,3,5$ $({\rm or}\,z\,)$.
We define $\Phi = e^{i P} S e^{i P}$ with $\langle S \rangle = v(z)$,
as advocated in ref.~\cite{Abidin:2009aj}.
In the chiral limit, the scalar vev spontaneously breaks
the chiral symmetry $SU(3)_L \otimes SU(3)_R$ down to $SU(3)_V$.
Under $SU(3)_V$, $S$ transforms as singlet,
and $P$ transforms as octet,
which is identified as pseudoscalar meson
after the Kaluza-Klein (KK) decomposition.
The AdS$_5$ space is characterized in the conformally flat metric
with a warp factor $a(z) \equiv L/z$,
\begin{equation}
ds^2 =  a^2 (z) ( dx^\mu dx_\mu - dz^2 ) \, .
\end{equation}
The scale $L$ is the curvature of the 5-dimensional AdS space.
In this model, the AdS$_5$ space is compactified such that
$L_0<z<L_1$, where $L_0\rightarrow 0$ is an ultra-violet (UV) cutoff
and $L_1$ is an infrared (IR) cutoff.
Solving the equation of motion (EOM) for $S$, we obtain~\cite{DaRold:2005zs}
\begin{eqnarray}
\langle S \rangle \equiv v(z) = {\rm diag} \{v_q, v_q, v_s\} = c_1 z + c_2 z^3
\end{eqnarray}
with the integration constants $c_{1,2}$,
\begin{eqnarray}
c_1 = \frac{M L_1^3 - \xi L_0^2}{L L_1 (L_1^2 - L_0^2)},
 \qquad c_2 = \frac{\xi - M L_1}{L L_1 (L_1^2 - L_0^2)}\; .
\end{eqnarray}
Here we adopted the following boundary conditions
\begin{eqnarray}
M = \frac{L}{L_0} v \bigg|_{L_0}, \qquad \xi = L v \bigg|_{L_1},
\end{eqnarray}
where $M={\rm diag}\{m_q, m_q, m_s\}$
is the $3\times 3$ current quark mass matrix,
which breaks chiral symmetry explicitly,
and the matrix $\xi={\rm diag}\{\xi_q, \xi_q, \xi_s\}$
is related to $\langle \bar{q} q \rangle$,
which  breaks chiral symmetry spontaneously.
As we can see, the isospin symmetry $SU(2)_V$
is still kept unbroken in the setup.

The vector and axial gauge bosons are defined by
\begin{eqnarray}
V_M &=& \frac{1}{\sqrt{2}} (L_M + R_M) \nonumber\\
A_M &=& \frac{1}{\sqrt{2}} (L_M - R_M)\, .
\end{eqnarray}

In order to cancel the mixing terms of $V_\mu$, $V_z$ and $S$,
the following gauge fixing term is added in the Lagrangian,
\begin{eqnarray}
{\cal L}^V_{\rm GF} &=& - \frac{M_5 a}{2 \xi_V} {\rm Tr} \bigg\{
   \partial_\mu V^\mu
   - \frac{\xi_V}{a} \bigg( \partial_5 (a V_z)
    - \frac{i}{\sqrt{2}} a^3 [S, v] \bigg) \bigg\}^2 \, .
\end{eqnarray}

In order to cancel the mixing terms of $A_\mu$, $A_z$ and $P$,
we add the following gauge fixing term,
\begin{eqnarray}
{\cal L}^A_{\rm GF} &=& - \frac{M_5 a}{2 \xi_A} {\rm Tr} \bigg\{
   \partial_\mu A^\mu
   - \frac{\xi_A}{a} \bigg( \partial_5 (a A_z)
    + \frac{1}{\sqrt{2}} a^3 (v^2 P + P v^2 + 2 v P v) \bigg) \bigg\}^2 \, .
\end{eqnarray}
In the unitary gauge, $\xi_A \to \infty$, we have the following relation
between $A_z$ and $P$,
\begin{eqnarray}
\label{AzP}
\partial_5 (a A_z)
  + \frac{1}{\sqrt{2}} a^3 (v^2 P + P v^2 + 2 v P v) = 0 \, .
\end{eqnarray}

The quadratic terms for vector, axial vector and pseudoscalar are given by,
\begin{eqnarray}
{\cal L}_V &=& \frac{M_5}{2} a {\rm Tr} \bigg(
  \partial_5 V_\mu \partial_5 V^\mu
  - \frac{1}{2} (\partial_\mu V_\nu - \partial_\nu V_\mu)
   (\partial^\mu V^\nu - \partial^\nu V^\mu)
  - \frac{1}{2} a^2 [V_\mu, v] [V^\mu, v] \bigg) \, , \nonumber\\ 
{\cal L}_A &=& \frac{M_5}{2} a {\rm Tr} \bigg(
  \partial_5 A_\mu \partial_5 A^\mu 
  - \frac{1}{2} (\partial_\mu A_\nu - \partial_\nu A_\mu)
   (\partial^\mu A^\nu - \partial^\nu A^\mu)
  + \frac{1}{2} a^2 \{A_\mu, v\} \{A^\mu, v\} \bigg) \, , \nonumber\\
{\cal L}_P &=& \frac{M_5}{2} a {\rm Tr} \bigg(
  \partial_\mu A_5 \partial^\mu A_5
  + a^2 \{\partial_\mu P, v\} \{\partial^\mu P, v\}
  - \frac{1}{2} a^2 \{A_5, v\} \{A_5, v\} \nonumber\\
&&\qquad\qquad  - a^2 \{\partial_z P, v\} \{\partial_z P, v\}
  - \sqrt{2} a^2 \{\partial_5 P, v\} \{A_5, v\}
  \bigg) \, .
\label{quardratic}
\end{eqnarray}

We also calculated $VVV$, $VAA$ and $VPP$ three-point interaction vertices,
\begin{eqnarray}
{\cal L}_{VVV} &=& - \frac{i}{2 \sqrt{2}} a M_5 {\rm Tr} \bigg(
  \{ \partial_\mu V_\nu, [V^\mu, V^\nu] \} \bigg) \, , \\
\label{VVV}
{\cal L}_{VAA} &=& - \frac{i}{2 \sqrt{2}} a M_5 {\rm Tr} \bigg(
  \{ \partial_\mu V_\nu, [A^\mu, A^\nu] \}
   + \{ \partial_\mu A_\nu - \partial_\nu A_\mu, [V^\mu, A^\nu] \} \bigg) \, ,\\
\label{VAA}
{\cal L}_{VPP} &=& \frac{i}{2 \sqrt{2}} a M_5 {\rm Tr} \bigg(
  \{ \partial_\mu A_5, [V^\mu, A_5] \} 
   + 2 a^2 \{ \partial_\mu P, v \} [V^\mu, \{P, v\}] \} \bigg) \, .
\label{VPP}
\end{eqnarray}
We have omitted the non-diagonal vector meson contributions
in $VPP$ interaction, eq.(\ref{VPP}),
which is irrelevant to our discussion in this work.

\section{Charged meson electromagnetic form factors}

In this section, we study the mass spectra, decay constants
and electromagnetic form factors of charged vector,
axial vector, pseudoscalar mesons.
The mass spectra, decay constants can be calculated
from the two-point correlation functions,
in the same way as done in ref.~\cite{Erlich:2005qh,DaRold:2005zs,Kim:2008xx}.
After KK decomposing the vector field as
$V_\mu(x,z) = \frac{1}{\sqrt{M_5 L}}
 \sum^{\infty}_{n=1} V^{(n)}_\mu (x) f^{(n)}_V(z)$,
we associate the first resonances
as $\rho$, $K^\ast$ vector mesons correspondingly
and omit the $(n)$ superscript index for them.
Explicitly, the $3 \times 3$ vector meson field $V_\mu$ can be written as,
\begin{eqnarray}
V_\mu = \frac{1}{\sqrt{2}} \left( \begin{array}{ccc}
   \frac{1}{\sqrt{2}} (\rho^0_\mu + \omega_\mu)  &
     \rho^+_\mu & K^{\ast +}_\mu \\
   \rho^-_\mu  &  - \frac{1}{\sqrt{2}} (\rho^0_\mu - \omega_\mu) &
     K^{\ast 0}_\mu \\
   K^{\ast -}_\mu  &  \bar{K}^{\ast 0}_\mu  &  \phi_\mu
 \end{array} \right) \, .
\end{eqnarray}
The same procedure and notation are adopted for axial and pseudoscalar mesons.

\subsection{Vector sector}

The EOM of vector $K^\ast$ meson can be written as
\begin{eqnarray}
\label{VEOM}
\partial^2_z f_{K^\ast} - \frac{1}{z} \partial_z f_{K^\ast} + m^2 f_{K^\ast}
  - \frac{1}{2} a^2 (v_s - v_q)^2 f_{K^\ast} = 0 \, ,
\end{eqnarray}
with $m$ as the mass of $K^\ast$ meson.
We impose the following boundary conditions to cancel the boundary terms,
\begin{eqnarray}
V_\mu|_{L_0} = 0, \qquad \partial_z V_\mu|_{L_1} = 0 \, .
\end{eqnarray}
The mass spectrum and wave function of $K^\ast$ meson
are obtained numerically with the normalization condition
$\int_{L_0}^{L_1} dz \frac{a}{L} f^{(m)}_{K^\ast}(z)  f^{(n)}_{K^\ast}(z) =
 \delta_{mn}$.

Before we study the electromagnetic form factors of $K^\ast$ meson,
we introduce the photon bulk-to-boundary propagator,
${\cal V}(q^2,z)$~\cite{Erlich:2006hq},
with the same EOM as $K^\ast$ but without the last term in eq.(\ref{VEOM}),
and the boundary conditions ${\cal V}(q^2,0)=1$,
$\partial_z{\cal V}(q^2,L_1)=0$.
We can derive the propagator as follow,
\begin{eqnarray}
{\cal V}(q^2,z) = \frac{\pi}{2 J_0(qL_1)} q z \bigg(
  J_1(q z) Y_0(qL_1) - Y_1(q z) J_0(qL_1) \bigg) \, .
\end{eqnarray}

The electromagnetic form factors of vector meson are defined
by the off-shell matrix element of electromagnetic current
between the same vector meson with different momentum and polarization.
Assuming P- and T- invariance,
the off-shell matrix element is parameterized
by three form factors $F_i$ ($i=1,2,3$),
\begin{eqnarray}
\label{Vffdef}
&&\hspace{-10mm} \langle K^{\ast+},p_2,\epsilon_2 | J_{\rm EM}^\mu(0) |
   K^{\ast+},p_1,\epsilon_1\rangle =
  (\epsilon_1 \cdot \epsilon_2^{\ast}) (p_1 + p_2)^{\mu} F_1(q^2)+ \nonumber\\
&& [\epsilon_2^{\ast\mu} (\epsilon_1 \cdot q) -
   \epsilon_1^\mu (\epsilon_2^\ast \cdot q)] [F_1(q^2) + F_2(q^{2})]
  + \frac{1}{p_2^2} (q \cdot \epsilon_1) (q \cdot \epsilon_2^\ast)
   (p_1 + p_2)^\mu F_3(q^2) \, ,
\end{eqnarray}
where $p_1, p_2$ and $\epsilon_1,\epsilon_2$ are
the momentum and polarization vector of initial and
final vector meson states, respectively.
$q = p_1 - p_2$ is the momentum of off-shell photon.

The form factors $F_i$ are related to the electric $F_E$,
magnetic $F_M$ and quadrupole $F_Q$ form factors by,
\begin{eqnarray}
F_E &=& F_1+ \frac{q^2}{6p_2^2} [F_2 - (1 - \frac{q^2}{4p_2^2}) F_{3}] \, ,
  \nonumber\\
F_M &=& F_1 + F_2 \, , \nonumber\\
F_Q &=& - F_2 + (1-\frac{q^2}{4p_2^2}) F_{3} \, .
\end{eqnarray}
The static observable, charge radius is defined as
\begin{eqnarray}
\langle r^2 \rangle \equiv - 6 \frac{\partial F_E(q^2)}{\partial q^2}
 \bigg|_{q^2=0} \, .
\end{eqnarray}

We can calculate the form factors $F_i(q^2)$ for $K^\ast$ as follows,
\begin{eqnarray}
F_1(q^2) = F_2(q^2) \equiv F(q^2), \qquad
F_3(q^2) = 0 \, , \nonumber
\end{eqnarray}
where the form factor $F(q^2)$ of charged $K^\ast$ meson is related to
the photon bulk-to-boundary propagator ${\cal V}(q^2,z)$
and $K^\ast$ meson wave function as,
\begin{eqnarray}
F(q^2) = \int_{L_0}^{L_1} dz \frac{a}{L} {\cal V}(q^2,z) f_{K^\ast}^2(z) \, ,
\end{eqnarray}
while the neutral $K^\ast$ meson form factor $F(q^2)$ vanishes.

\subsection{Axial vector sector}

The EOM of axial vector $K_1$ meson can be written as
\begin{eqnarray}
\label{AEOM}
\partial^2_z f_{K_1} - \frac{1}{z} \partial_z f_{K_1} + m^2 f_{K_1}
  - \frac{1}{2} a^2 (v_s + v_q)^2 f_{K_1} = 0 \, .
\end{eqnarray}
With similar boundary conditions as the vector $K^\ast$ meson,
we can derive the mass spectrum of $K_1$ meson.
The electromagnetic form factor of $K_1$ meson is also similar
to the $K^\ast$ meson.
We can calculate the form factor $F(q^2)$ of charged $K_1$ as,
\begin{eqnarray}
F(q^2) = \int_{L_0}^{L_1} dz \frac{a}{L} {\cal V}(q^2,z) f_{K_1}^2(z) \, ,
\end{eqnarray}
while the neutral counterpart vanishes.

\subsection{Pseudoscalar sector}

The EOM of pseudoscalar $K$ meson can be written as
\begin{eqnarray}
\label{PEOM}
&& \partial_z f_{P_K} + \frac{1}{\sqrt{2}}
  (1 - \frac{m^2}{a^2 (v_s + v_q)^2}) f_{A_K} = 0 \nonumber\\
&& \partial^2_z f_{P_K} +
   \partial_z f_{P_K} ( -3 a + 2 \partial_z \ln (v_s + v_q) )
  + m^2 f_{P_K} - \frac{1}{2} a^2 (v_s + v_q)^2 f_{P_K} \nonumber\\
&& \qquad  + \frac{1}{\sqrt{2}} f_{A_K}
    ( -2 a + 2 \partial_z \ln (v_s + v_q) ) = 0 \, ,
\end{eqnarray}
together with the constraint eq.(\ref{AzP}),
we can solve the differential equations
with the following boundary conditions
in order to cancel the boundary terms,
\begin{eqnarray}
f_{P_K}|_{L_0} = 0, \quad f_{A_K}|_{L_1} = 0, \quad
 \partial_z f_{P_K}|_{L_1} = 0 \, . \nonumber
\end{eqnarray}
The kaon wave function can be normalized with the normalization condition,
\begin{eqnarray}
\int_{L_0}^{L_1} dz \bigg( a f_{A_K}^2 + a^3 (v_s + v_q)^2 f_{P_K}^2
 \bigg) = 0 \, . \nonumber
\end{eqnarray}

The electromagnetic form factor of pseudoscalar meson is defined in
\begin{eqnarray}
\langle K^+(p_2) | J^\mu_{\rm EM}(0) | K^+(p_1) \rangle =
 F(q^2) (p_1 + p_2)^\mu,
\end{eqnarray}
with $q = p_1 - p_2$ as the momentum transfer of off-shell photon.

With the help of photon bulk-to-boundary propagator,
the form factor of charged kaon can be written as
\begin{eqnarray}
F(q^2) = \int_{L_0}^{L_1} dz {\cal V}(q^2,z)
  \bigg( a f_{A_K}^2 + a^3 (v_s + v_q)^2 f_{P_K}^2 \bigg) \, ,
\end{eqnarray}
while the form factor of neutral kaon vanishes.

\subsection{Numerical results}

In this subsection, we present the numerical results of
the mass spectra, decay constants
and electromagnetic form factors of charged vector,
axial vector, pseudoscalar mesons.
We first fit the three parameters $L_1$, $\xi_q$ and $m_q$
from $m_\rho$, $m_{a_1}$ and $m_\pi$ in the case of $SU(2)$ only.
We find the following best fitting parameters:
$L_1=3.1$ ${\rm GeV}^{-1}$, $m_q=0.01$ GeV and $\xi_q=4.0$,
which are in agreement with the parameters in ref.~\cite{DaRold:2005zs}.
Proceeding to the $SU(3)$ case, we fit $\xi_s$ and $m_s$
from $K^\ast$ and $K$ masses in the case A
with $\xi_s = 6.8$, $m_s = 0.225$ GeV as best fitting parameters,
and fit them from $K_1$ and $K$ masses in the case B,
with $\xi_s = 3.4$, $m_s = 0.244$ GeV.
Our main numerical results are summarized
in Table {\bf 1}-{\bf 3},
and the electric form factors are plotted in Figure {\bf 1}-{\bf 3}.

\begin{table}[!htb]
\begin{center}
\begin{tabular}{|c|c|c|c|}
\hline
  &  $\rho$  &  $a_1$  &  $\pi$  \\
\hline
mass (GeV)  & [$0.776$] $(0.775)$ & [$1.267$] $(1.230)$ & [$0.136$] $(0.135)$ \\
\hline
decay const. (GeV)  & $0.140$ & $0.162$ & $0.087$ $(0.092)$ \\
\hline
$\langle r^2 \rangle$ (fm$^2$) & $0.527$ & $0.416$ & $0.349$ $(0.452)$ \\
\hline
\end{tabular}
\caption{\label{casesu2}
Mass spectra, decay constants and charge radii
of charged $\rho$, $a_1$ and $\pi$.
Fit $L_1$, $m_q$, $\xi_q$ parameters
with the masses of $\rho$, $a_1$ and $\pi$ in the case of $SU(2)$ only.
The fitted parameters are $L_1=3.1$ ${\rm GeV}^{-1}$,
$m_q=0.01$ GeV and $\xi_q=4.0$.
The inputs are shown in the square brackets,
and experiment data in the round brackets.
}
\end{center}
\end{table}

\begin{table}[!htb]
\begin{center}
\begin{tabular}{|c|c|c|c|}
\hline
  &  $K^\ast$  &  $K_1$  &  $K$  \\
\hline
mass (GeV)  & [$0.892$] $(0.892)$ & $1.500$ $(1.272)$ & [$0.494$] $(0.494)$  \\
\hline
decay const. (GeV)  & $0.135$ & $0.188$ & $0.120$ $(0.110)$ \\
\hline
$\langle r^2 \rangle$ (fm$^2$) & $0.504$ & $0.362$ & $0.305$ $(0.314)$ \\
\hline
\end{tabular}
\caption{\label{caseA}
Case A: Mass spectra, decay constants and charge radii
of charged $K^\ast$, $K_1$ and $K$.
Fit $\xi_s$, $m_s$ with the masses of $K^\ast$ and $K$
with strange sector taken into account.
The fitted parameters are $\xi_s = 6.8$ and $m_s = 0.225$ GeV.
}
\end{center}
\end{table}

\begin{table}[!htb]
\begin{center}
\begin{tabular}{|c|c|c|c|}
\hline
  &  $K^\ast$  &  $K_1$  &  $K$  \\
\hline
mass (GeV)  & $0.780$ $(0.892)$ & [$1.270$] $(1.272)$ & [$0.493$] $(0.494)$  \\
\hline
decay const. (GeV)  & $0.138$ & $0.163$ & $0.111$ $(0.110)$ \\
\hline
$\langle r^2 \rangle$ (fm$^2$) & $0.527$ & $0.425$ & $0.344$ $(0.314)$ \\
\hline
\end{tabular}
\caption{\label{caseB}
Case B: Mass spectra, decay constants and charge radii
of charged $K^\ast$, $K_1$ and $K$.
Fit $\xi_s$, $m_s$ with the masses of $K_1$ and $K$
with strange sector taken into account.
The fitted parameters are $\xi_s = 3.4$ and $m_s = 0.244$ GeV.
}
\end{center}
\end{table}

In case A, we fit $\xi_s$ and $m_s$ from $K^\ast$ and $K$ masses.
The $K_1$ mass $1.5$ GeV is much larger than the $K_1(1270)$ ground state,
and even larger than the 1st excitation state $K_1(1400)$.
While in case B, we fit $\xi_s$ and $m_s$ from $K_1$ and $K$ masses.
The $K^\ast$ mass $0.78$ GeV is smaller than $K^\ast(892)$
and close to $\rho(770)$ mass.
The $K$ decay constant is in good agreement with the experimental value
in case B, while it is a little larger in case A.
The calculated values of $K^\ast$ decay constant are similar
in both cases, for $K_1$ decay constant,
the value in case A is larger than the one in case B.

The static observable, charge radii of charged pion and kaon
have been measured by experiments.
The pion form factors have been determined by experiments
at CERN~\cite{Amendolia:1983di}, DESY~\cite{Brauel:1977ra,Ackermann:1977rp}
and JLab~\cite{Tadevosyan:2007yd,Horn:2006tm},
and, the charge radius of charged kaon
has been measured by UCLA+~\cite{Dally:1980dj}
and CERN NA7 Collaborations~\cite{Amendolia:1986ui},
with the average in Table {\bf 1}-{\bf 3}.
The charge radius of charged pion is smaller than the experimental value,
as obtained in other calculations
in the hard-wall holographic QCD models~\cite{Brodsky:2007hb,Kwee:2007dd,
Grigoryan:2007wn,Kwee:2007nq,Kwee:2008zt,Kim:2008xx}.
The charge radius of charged kaon is in good agreement with
the experimental value in both cases,
even though it is a little larger in case B.
There are no experiment data on the charge radii of
both vector and axial vector mesons.
The charged rho charge radius agrees with previous calculations
in the framework of holographic QCD~\cite{Grigoryan:2007vg,
Grigoryan:2007my,BallonBayona:2009ar},
and a recent DSE calculation~\cite{Bhagwat:2006pu}.
For charged $K^\ast$, our results are only a little bit larger than
the DSE calculation~\cite{Bhagwat:2006pu}
with $\langle r^2 \rangle = 0.43$ ${\rm fm}^2$.
We also present our result of the charge radii of
charged axial vector $a_1$ and $K_1$ mesons in Table {\bf 1}-{\bf 3}.

The electric form factors of charged mesons are plotted
in Figure {\bf 1}-{\bf 2} in both space-like and time-like regions.
In case A, the form factors of strange mesons are systematically larger
than those of un-flavored mesons with the same quantum numbers
correspondingly in the space-like region,
and smaller in the time-like region.
The form factors of strange mesons are quite close to
those of light un-flavored mesons in case B,
because the chiral condensate $\xi_s$ is close to $\xi_q$.
In Figure {\bf 3}, we find agreement with
the experiment data from UCLA+~\cite{Dally:1980dj},
and CERN~\cite{Amendolia:1986ui} in both cases,
when we zoom in the charged kaon form factor
as a function of momentum square,
although with large experiment uncertainties.

\begin{figure}[!htb]
\centerline{\includegraphics[width=16cm] {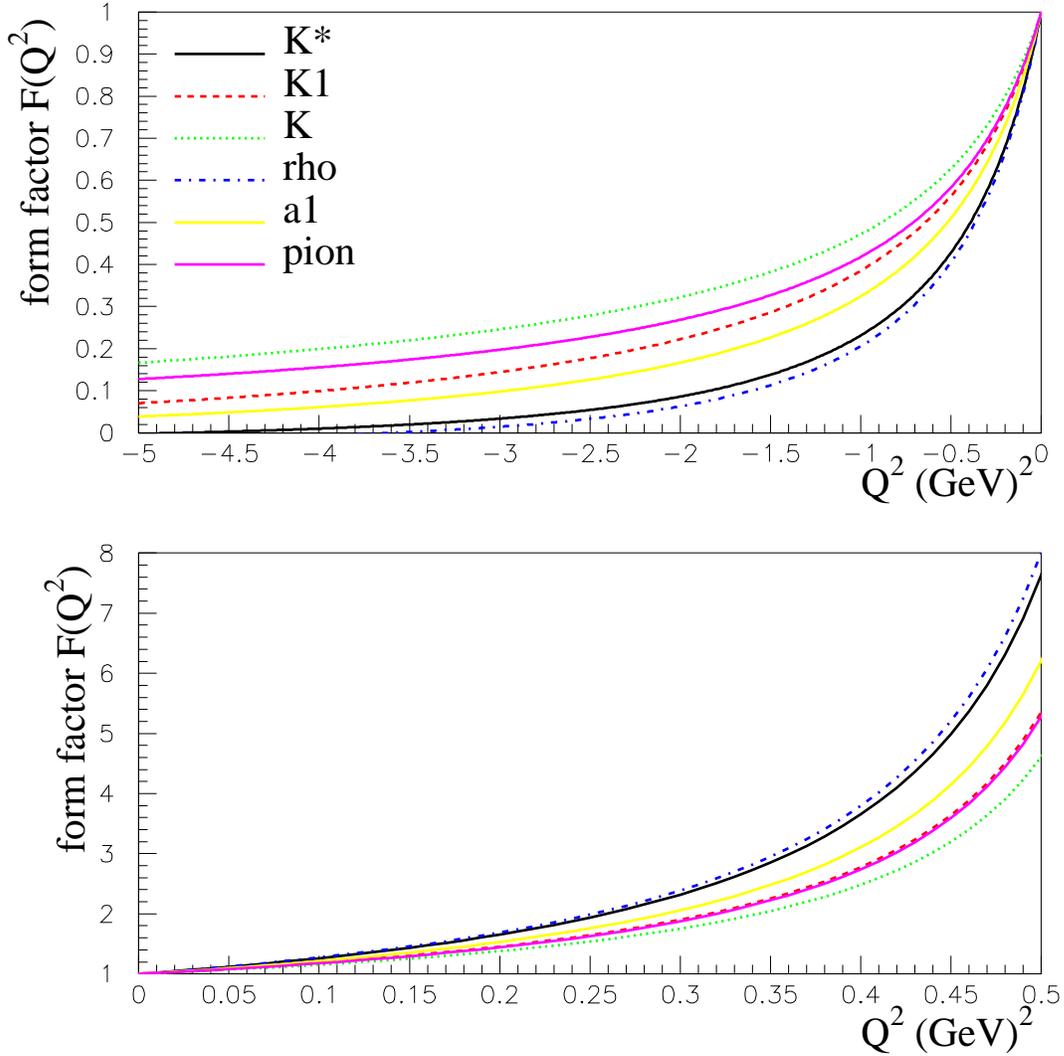}}
\caption{ \label{formfactorA}
Case A: Electric form factor $F(Q^2)$ as a function of $Q^2$
for charged mesons in both space-like and time-like regions.
}
\end{figure}

\begin{figure}[!htb]
\centerline{\includegraphics[width=16cm] {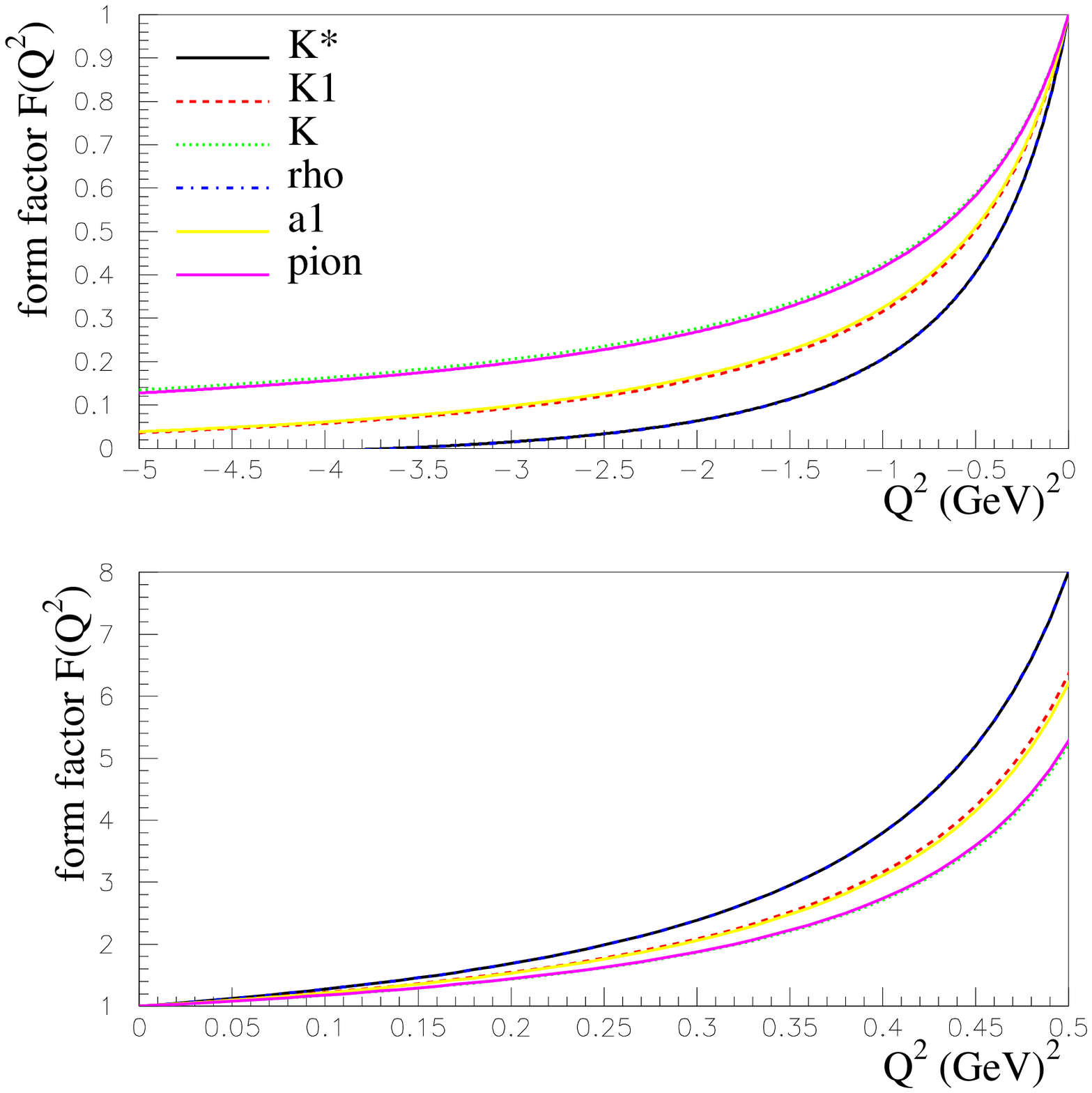}}
\caption{ \label{formfactorB}
Case B: Electric form factor $F(Q^2)$ as a function of $Q^2$
for charged mesons in both space-like and time-like regions.
}
\end{figure}

\begin{figure}[!htb]
\centerline{\includegraphics[width=10cm] {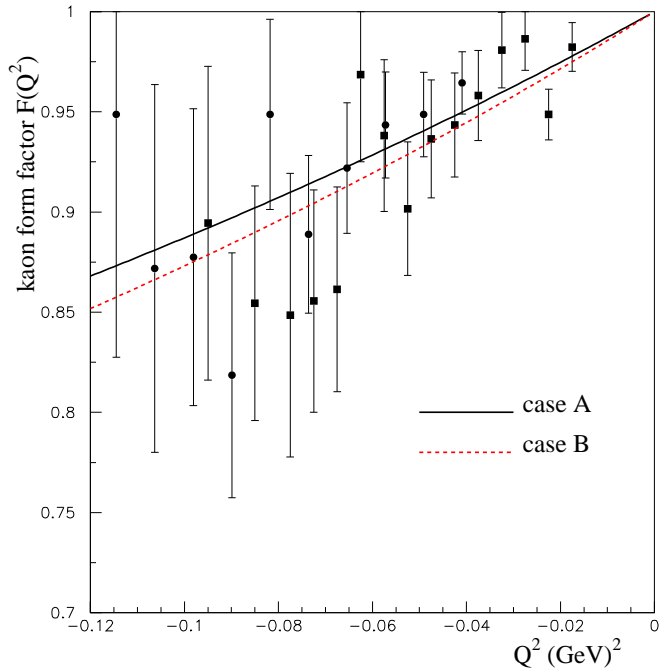}}
\caption{ \label{kaonformfactor}
The electric form factor $F(Q^2)$ of charged kaon as a function of $Q^2$.
The black circles are data from UCLA+~\cite{Dally:1980dj},
and the black square from CERN~\cite{Amendolia:1986ui}.
}
\end{figure}

\section{Conclusions}

We have studied the mass spectra, decay constants,
charge radii and electromagnetic form factors of
strange vector, axial vector and pseudoscalar mesons
in a holographic QCD model.
We find the decay constant and charge radius
of charged kaon agree with the experiment,
while those of charged pion are a little bit smaller than
the experimental value, as obtained in other calculations
in the hard-wall holographic QCD models.
The charge radii of charged rho and $K^\ast$ quantitatively
agree with a recent DSE calculation.
We also present the electric form factors of vector, axial vector
and pseudoscalar mesons in both space-like and time-like regions.
The electric form factor of charged kaon is in agreement with
the experiment data.

\acknowledgments

The authors would like to acknowledge
the collaboration with Youngman Kim and P. Ko in the initial stage.
This work was supported in part by the National Natural Science Foundation of
China under grant No. 10805018.

\end{document}